\documentclass[11pt,twoside]{article}

\usepackage{asp2006}
\usepackage{epsf}
\usepackage{psfig}
\usepackage{lscape}

\begin{document}
\title{Resolved Stellar Populations Constituting Extended UV Disks (XUV-disks) in Nearby Galaxies}   
\author{D. A. Thilker\altaffilmark{1}, L. Bianchi\altaffilmark{1}, G. Meurer\altaffilmark{1}, A. Gil de Paz\altaffilmark{2}, S. Boissier\altaffilmark{3}, B. Madore\altaffilmark{4}, A. Ferguson\altaffilmark{5}, S. Hameed\altaffilmark{6}, S. Neff\altaffilmark{7}, C. D. Martin\altaffilmark{8}, R. M. Rich\altaffilmark{9}, D. Schiminovich\altaffilmark{10}, M. Seibert\altaffilmark{4}, and T. Wyder\altaffilmark{8}}   
\altaffiltext{1}{Center for Astrophysical Sciences, JHU, 3400 N. Charles St., Baltimore, MD 21218}
\altaffiltext{2}{Departamento de Astrof\'{\i}sica, Universidad Complutense de Madrid, Madrid 28040, Spain}
\altaffiltext{3}{Laboratoire d'Astrophys. de Marseille, BP 8, Traverse du Siphon, 13376 Marseille Cedex 12, France}
\altaffiltext{4}{The Observatories, Carnegie Inst. of Washington, 813 Santa Barbara St., Pasadena, CA 91101}
\altaffiltext{5}{Institute for Astronomy, Univ. of Edinburgh, Royal Observatory Edinburgh, Edinburgh, UK}
\altaffiltext{6}{Five College Astronomy Department, Hampshire College, Amherst, MA 01003}
\altaffiltext{7}{Laboratory for Astronomy and Solar Physics, NASA GSFC, Greenbelt, MD 20771}
\altaffiltext{8}{Caltech, MC 40547, 1200 E. California Boulevard, Pasadena, CA 91125}
\altaffiltext{9}{Department of Physics and Astronomy, University of California, Los Angeles, CA 90095}
\altaffiltext{10}{Department of Astronomy, Columbia University, New York, NY 10027}

\begin{abstract} 
We describe HST imaging of recent star formation complexes located in
the extended UV disk (XUV-disk) component of NGC 5236 (M 83), NGC 5055
(M 63), and NGC 2090.  Photometry in four FUV--visible bands
permits us to constrain the type of resolved stars and
effective age of clusters, in addition to extinction.  The preliminary
results given herein focus on CMD analysis and clustering properties in this
unique star-forming environment.  

\end{abstract}


\section{Overview and Preliminary Results}   
A decade ago, deep H$\alpha$ observations indicated that some disk
galaxies can support limited star formation at their extreme outer
edge (e.g. Ferguson et al. 1998).  GALEX imaging then surprisingly
revealed that M 83 (Thilker et al. 2005) and NGC 4625 (Gil de Paz et
al. 2005) have extended UV disks (XUV-disks) unapparent in the
distribution of HII regions. We have since demonstrated that outer
disk SF activity is commonplace, with $\sim$ 1/3 of nearby S0-Sm
galaxies having discernible XUV-disk structure (Thilker et al. 2007).
For detailed information, see the review by Gil de Paz (this volume) or
Thilker et al. (2007).

The relative lack of HII regions compared to UV clumps in the low SFR
outer disk has been largely explained as a stochastic effect, tied to
the very limited HII region lifetime compared to the time-scale for UV
production (Boissier et al. 2007). However, alternative contributing
factors (top-light IMF, low density ISM) have yet to be
ruled out and motivate our HST analysis.

HST ACS UV--visible imaging of eight XUV-disk fields was obtained for
M83. Single locations in each of NGC 5055 (Fig. 1) and NGC 2090 are also being
studied.  We observed in four band-passes (F150LP, F435W,
F606W, and F814W) using the WFC and SBC.  Optical observations of NGC
2090 were obtained using WFPC2 (after the failure of ACS/WFC).

\begin{figure}
\plottwo{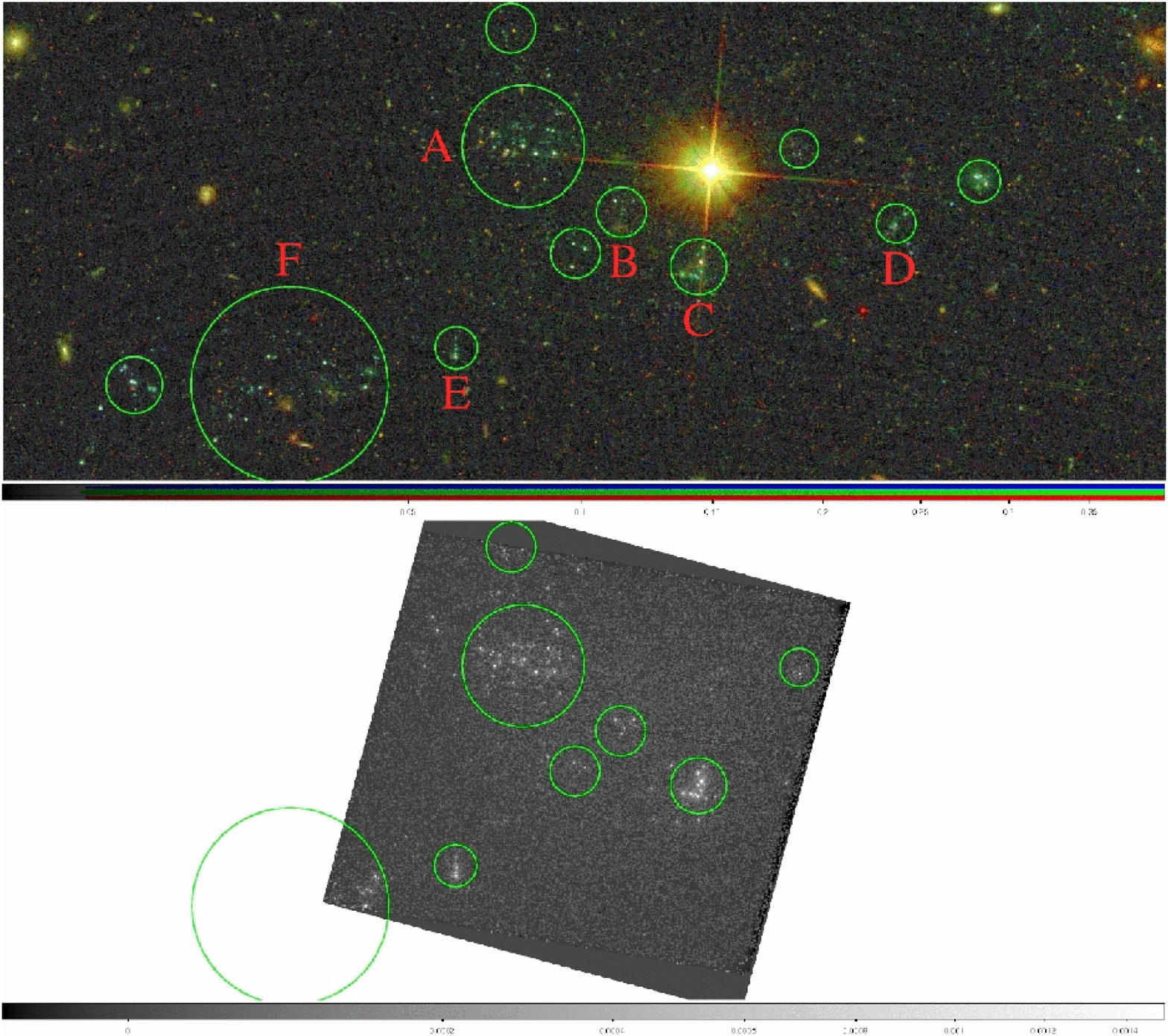}{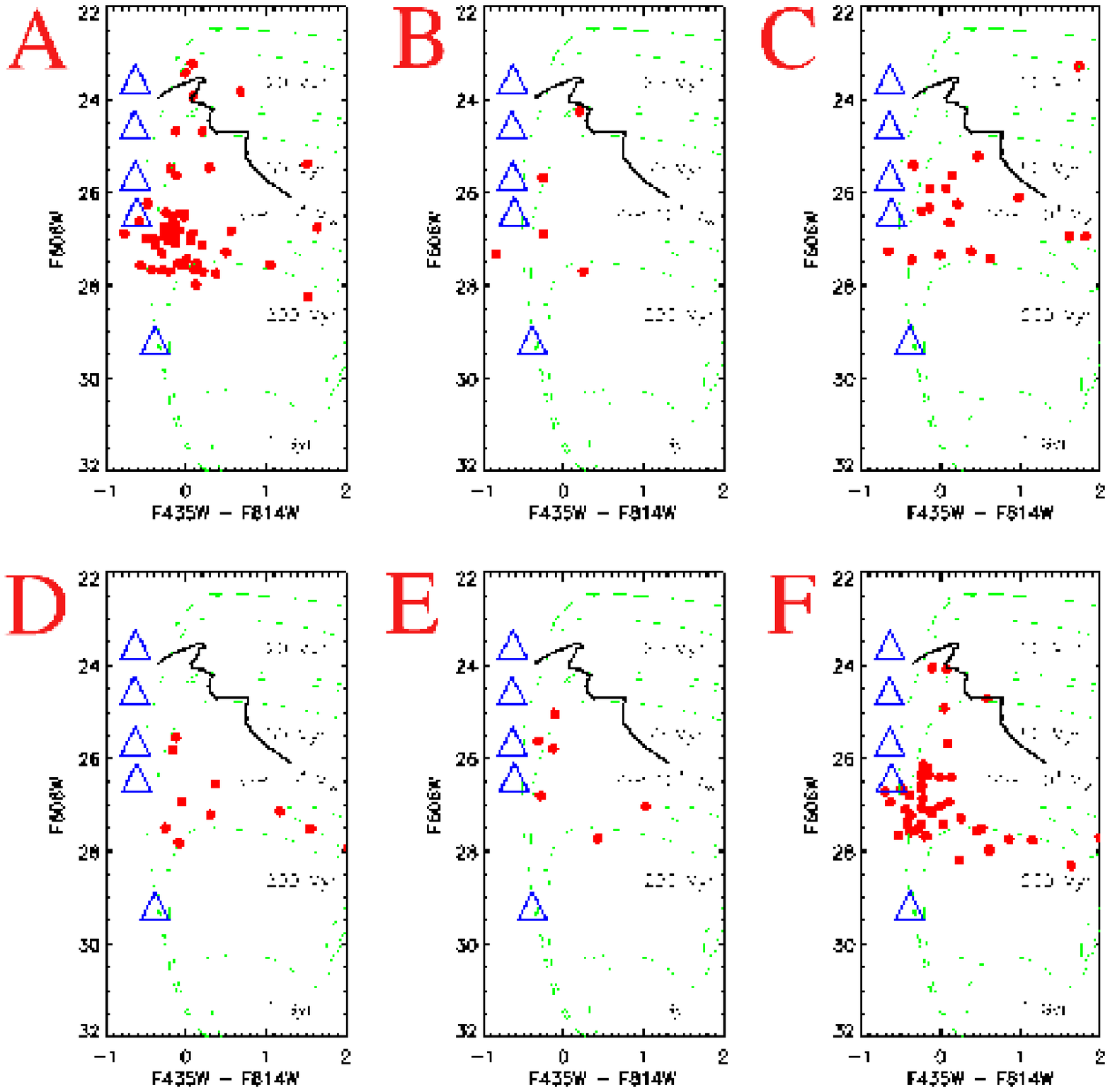}
\caption{(Left) HST imaging of NGC~5055's XUV-disk.  F814W, F606W, and F435W bands are shown above in RGB repesentation, while the lower panel displays UV emission. We mark several SF complexes with circles. (Right) Foreground ext.-corrected CMDs for six of the
circled complexes.  We show the ZAMS position for
stars of 5, 15, 20, 30, and 40 M$_\odot$ (blue triangles).  Dashed green lines are Padova
isochrones for populations of age 20, 50, 200 Myr and 1 Gyr.  The thick
black line shows the evolution of an SSP burst totaling 10$^3$ M$_\odot$ from 5 Myr to 1 Gyr.}
\end{figure}

HST resolves the XUV-disk sources into loosely clustered complexes of
individual stars.  These complexes, likely evolved OB associations,
are low mass ($< 10^3$ M$_\odot$), intermediate age structures.  Only
very few HST detections are consistent with being zero-age upper-MS
stars having mass $>$ 15 M$_\odot$ (Fig. 1). H$\alpha$ emission is
detected from complexes in which they are found.  Observed association
sizes vary from 100 pc to $\sim$ 500 pc with significant internal
sub-clustering.  The largest groupings may be blended
associations.  CMDs (Fig. 1) suggest multiple
generations within larger complexes (up to age of $\sim$ 200 Myr).


\acknowledgements 
This work was supported by grants HST-GO-10608.01-A and HST-GO-10904.01-A.



\begin{thebibliography}{}
\bibitem[Boissier et al. 2007]{Boissier et al. 2007}Boissier, S., et al. 2007, ApJS, 173, 524
\bibitem[Ferguson et al. 1998]{Ferguson et al. 1998}Ferguson, A., et al. 1998, ApJ, 506, L19
\bibitem[Gil de Paz et al. 2005]{Gil de Paz et al. 2005}Gil de Paz, A., et al. 2005, ApJ, 627, L29
\bibitem[Thilker et al. 2005]{Thilker et al. 2005}Thilker, D. A., et al. 2005, ApJ, 619, L79
\bibitem[Thilker et al. 2007]{Thilker et al. 2007}Thilker, D. A., et al. 2007, ApJS, 173, 538
\end{thebibliography}
\end{document}